\documentclass[12pt]{article}%
\usepackage{amsmath,latexsym}
\usepackage{graphicx}
\usepackage{amsmath}
\usepackage{amsfonts}
\usepackage{amssymb}%
\begin{document}

\title{{Wormholes admitting conformal Killing vectors and
supported by generalized Chaplygin gas }}
   \author{
  Peter K.F. Kuhfittig\\  \footnote{kuhfitti@msoe.edu}
 \small Department of Mathematics, Milwaukee School of
Engineering,\\
\small Milwaukee, Wisconsin 53202-3109, USA}

\date{}
 \maketitle

\begin{abstract}\noindent
When Morris and Thorne first proposed that
traversable wormholes may be actual physical
objects, they concentrated on the geometry by
specifying the shape and redshift functions.
This mathematical approach necessarily raises
questions regarding the determination of the
required stress-energy tensor.  This paper
discusses a natural way to obtain a complete
wormhole solution by assuming that the
wormhole (1) is supported by generalized
Chaplygin gas and (2) admits conformal
Killing vectors.
\\

\noindent
%PAC numbers: 04.20.Jb, 04.20.-q, 04.20.Gz
\end{abstract}

\section{Introduction}\label{E:introduction}

Wormholes are handles or tunnels in spacetime
connecting different regions of our Universe or
different universes altogether.  That wormholes
could be actual structures suitable for
interstellar travel was first proposed by Morris
and Thorne \cite{MT88}.  Such a wormhole could
be described by the static spherically symmetric
line element \cite{MT88}
\begin{equation}\label{E:line1}
ds^{2}=-e^{2\Phi(r)}dt^{2}+\frac{dr^2}{1-b(r)/r}
+r^{2}(d\theta^{2}+\text{sin}^{2}\theta\,
d\phi^{2}),
\end{equation}
using units in which $c=G=1$.  Here $\Phi=
\Phi(r)$ is called the \emph{redshift function},
which must be everywhere finite to avoid an
event horizon.  The function $b=b(r)$ helps
determine the spatial shape of the wormhole
and is therefore called the \emph{shape
function}.  The spherical surface $r=r_0$ is
the \emph{throat} of the wormhole and must
satisfy the following conditions: $b(r_0)=r_0$,
$b(r)<r$ for $r>r_0$, and $b'(r_0)<1$, now
usually called the \emph{flare-out condition}.
This condition refers to the flaring out of
the embedding diagram pictured in Ref.
\cite{MT88}.  The flare-out condition can
only be satisfied by violating the null
energy condition.

The Einstein field equations in the orthonormal
frame, $G_{\hat{\mu}\hat{\nu}}=8\pi
T_{\hat{\mu}\hat{\nu}}$, yield the following
simple interpretation for the components of the
stress-energy tensor: $T_{\hat{t}\hat{t}}=
\rho(r)$, the energy density,
$T_{\hat{r}\hat{r}}=p_r$, the radial pressure,
and $T_{\hat{\theta}\hat{\theta}}=
T_{\hat{\phi}\hat{\phi}}=p_t$, the lateral
pressure.  For the theoretical construction of
the wormhole, Morris and Thorne then proposed
the following strategy: retain complete control
over the geometry by specifying the functions
$b(r)$ and $\Phi(r)$ to obtain the desired
properties of the wormhole.  The problem with
this strategy is that it relies on the
engineering team to manufacture or search for
those materials or fields that yield the
required stress-energy tensor.

This purely geometric approach can be
supplemented by an appropriate equation of
state, an example of which is $p=\omega\rho$,
$\omega<-1$, representing phantom dark energy
\cite{fL05, sS05,pK06,pG06}.  The energy
density may also be known for physical
reasons, as in the case of the
Navarro-Frenk-White density profile for dark
matter \cite{NFW90,RKRI14}
\begin{equation}
  \rho(r)=\frac{\rho_s}{\frac{r}{r_s}
  \left(1+\frac{r}{r_s}\right)^2},
\end{equation}
where $r_s$ is the characteristic scale
radius and $\rho_s$ is the corresponding
density.

In this paper we consider another equation
of state, $p=-A/\rho$, $A>0$, representing
Chaplygin gas or, more generally,
$p=-A/\rho^{\alpha},$ $0<\alpha\le 1$,
called generalized Chaplygin gas
\cite{KMP01,BBS02,fL06,pK08,eE09}.
Cosmologists became interested in this model
when it was shown that Chaplygin gas is a
candidate for unifying dark matter and dark
energy.  To see this, consider the energy
conservation equation $\dot{\rho}=-3\dot{a}
(\rho+p)/a$ in a flat FLRW spacetime and
then substitute $p=-A/\rho^{\alpha}$.  The
result is
\begin{equation}
  \rho=\left(A+\frac{B}{a^{3(\alpha +1)}}
  \right)^{1/(\alpha +1)},
\end{equation}
where $B$ is an integration constant.  It
is seen that $\rho\sim a^{-3}$ at early
times, implying that $\rho$ behaves like
matter, while in later times it behaves
like a cosmological constant
($\rho\equiv\text{constant})$.

 In a cosmological setting one would
 normally assume a homogeneous distribution
 of matter.  It was pointed out in Ref.
 \cite{BP05}, however, that the generalized
 Chaplygin equation of state is that of a
 polytropic gas with a negative polytropic
 index.  Thus inhomogeneous structures may
 arise from a density fluctuation in the
 cosmological background.  Also, a
 Born-Infeld phantom gravastar has been
 constructed by replacing the interior
 de Sitter solution with the Chaplygin
 gas equation of state \cite{BTV06}.  A
 similar problem arises with another type
 of dark energy, phantom dark energy.
 Here Sushkov and Kim \cite{SK04} have
 shown that away from the throat the
 radial and transverse pressures converge
 fairly quickly.  Accordingly, we will
 follow Refs. \cite{fL06,pK08} and assume
 that the equation of state is
 \begin{equation}\label{E:Chaplygin}
    p_r=-\frac{A}{\rho^{\alpha}}.
 \end{equation}
The tangential pressure can be determined
from the Einstein field equations.

To obtain a complete wormhole solution,
we are going to make the additional
assumption that our spacetime admits a
one-parameter group of conformal motions,
i.e., conformal Killing vectors, discussed
next.
%END OF SECTION

\section{Conformal Killing vectors}

As noted above, we assume that our spacetime
admits a one-parameter group of conformal
motions, which are motions along which the
metric tensor of a spacetime remains
invariant up to a scale factor.  In other
words, there exists a set of conformal
Killing vectors such that
\begin{equation}\label{E:Lie}
   \mathcal{L_{\xi}}g_{\mu\nu}=g_{\eta\nu}\,\xi^{\eta}
   _{\phantom{A};\mu}+g_{\mu\eta}\,\xi^{\eta}_{\phantom{A};
   \nu}=\psi(r)\,g_{\mu\nu},
\end{equation}
where the left-hand side is the Lie derivative of the
metric tensor and $\psi(r)$ is the conformal factor.  The
vector $\xi$ characterizes the conformal symmetry since the
metric tensor $g_{\mu\nu}$ is conformally mapped into
itself along $\xi$.  The assumption of conformal symmetry
has led to numerous new solutions, as well as new
geometric and kinematical insights
\cite{HPa, HPb, MM96, MS93, Ray08, fR10, fR12}.

Given a noncommutative-geometry background, exact
solutions of traversable wormholes admitting conformal
motions are discussed in Ref. \cite{R2K3}.  Two earlier
studies assumed a \emph{non-static} conformal symmetry
\cite{BHL07, BHL08}.

As in Ref. \cite{R2K3}, we use the following form of
the metric to discuss conformal symmetry:
\begin{equation}\label{E:line2}
   ds^2=- e^{\nu(r)} dt^2+e^{\lambda(r)} dr^2
   +r^2( d\theta^2+\text{sin}^2\theta\, d\phi^2).
\end{equation}
The Einstein field equations are:

\begin{equation}\label{E:Einstein1}
e^{-\lambda}
\left[\frac{\lambda^\prime}{r} - \frac{1}{r^2}
\right]+\frac{1}{r^2}= 8\pi \rho,
\end{equation}

\begin{equation}\label{E:Einstein2}
e^{-\lambda}
\left[\frac{1}{r^2}+\frac{\nu^\prime}{r}\right]-\frac{1}{r^2}=
8\pi p_r,
\end{equation}

\noindent and

\begin{equation}\label{E:Einstein3}
\frac{1}{2} e^{-\lambda} \left[\frac{1}{2}(\nu^\prime)^2+
\nu^{\prime\prime} -\frac{1}{2}\lambda^\prime\nu^\prime +
\frac{1}{r}({\nu^\prime- \lambda^\prime})\right] =8\pi p_t.
\end{equation}
Eq. (\ref{E:Einstein3}) can actually be obtained from the
conservation of the stress-energy tensor, i.e.,
$T^{\mu\nu}_{\phantom{\mu\nu};\nu}=0$.  So we need to use
only Eqs. (\ref{E:Einstein1}) and (\ref{E:Einstein2}).

To discuss the assumption of conformal symmetry in
Eq. (\ref{E:Lie}), we follow Herrera and
Ponce de Le\'{o}n \cite{HPa} and restrict the vector
field by requiring that $\xi^{\alpha}U_{\alpha}=0$,
where $U_{\alpha}$ is the four-velocity of the perfect
fluid distribution.  The assumption of spherical
symmetry then yields $\xi^0=\xi^2=\xi^3=0$ \cite{HPa}.
Eq. (\ref {E:Lie}) now produces the following results:
\begin{equation}\label{E:sol1}
    \xi^1 \nu^\prime =\psi,
\end{equation}
\begin{equation}\label{E:sol2}
   \xi^1  = \frac{\psi r}{2},
\end{equation}
and
\begin{equation}\label{E:sol3}
  \xi^1 \lambda ^\prime+2\,\xi^1 _{\phantom{1},1}=\psi.
\end{equation}
These equations, in turn, yield
\begin{equation} \label{E:gtt}
   e^\nu  =C r^2
\end{equation}
and
\begin{equation}\label{E:grr}
   e^\lambda  = \left(\frac {a} {\psi}\right)^2,
\end{equation}
where $C$ and $a$ are integration constants.  In
order to make use of these equations, it is convenient
to write Eqs. (\ref{E:Einstein1}) and (\ref{E:Einstein2})
in the following forms:
\begin{equation}\label{E:E1}
\frac{1}{r^2}\left(1 - \frac{\psi^2}{a^2}
\right)-\frac{2\psi\psi^\prime}{a^2r}= 8\pi \rho
\end{equation}
and
\begin{equation}\label{E:E2}
\frac{1}{r^2}\left( \frac{3\psi^2}{a^2}-1
\right)= 8\pi p_r.
\end{equation}

%END OF SECTION

\section{The solution}

To obtain a wormhole solution, we start with
the equation of state (\ref{E:Chaplygin}),
$p_r=-A/\rho^{\alpha}$, and substitute
Eqs. (\ref{E:E1}) and (\ref{E:E2}) to obtain
\begin{equation}\label{E:diff1}
   \frac{1}{8\pi}\frac{1}{r^2}\left(
   \frac{3\psi^2}{a^2}-1\right)=
   -\frac{A}{\left\{\frac{1}{8\pi}\left[
   \frac{1}{r^2}\left(1-\frac{\psi^2}{a^2}
   \right)-\frac{2\psi\psi'}{a^2r}
   \right]\right\}^{\alpha}}.
\end{equation}
This equation can be put into a more
transparent form by noting that
$2\psi\psi'=(\psi^2)'$ and eliminating the
negative sign:
\begin{equation}\label{E:diff2}
  \frac{1}{8\pi}\frac{1}{r^2}
  \left(1-\frac{3\psi^2}{a^2}\right)=
  \frac{A}{\left\{\frac{1}{8\pi}\left[
  \frac{1}{r^2}\left(1-\frac{\psi^2}{a^2}
  \right)-\frac{(\psi^2)'}{a^2r}
  \right]\right\}^{\alpha}}.
\end{equation}
Now observe that since $A$ and
$\rho^{\alpha}$ are positive, the left
side is also positive, allowing us to
raise each side to the power $1/\alpha$.
We can thereby obtain
\begin{equation}\label{E:diff3}
   (\psi^2)'=\frac{1}{r}\left[a^2-\psi^2
   -\left(\frac{Aa^{2(\alpha +1)}
   (8\pi)^{\alpha +1}r^{2(\alpha +1)}}{a^2-3\psi^2}
   \right)^{1/\alpha}\right].
\end{equation}

Before considering a possible solution,
observe that by comparing Eqs. (\ref{E:line1})
and (\ref{E:line2}), we have in view of
Eq. (\ref{E:grr}),
\begin{equation}\label{E:shape1}
  b(r)=r(1-e^{-\lambda})=
  r\left(1-\frac{\psi^2}{a^2}\right).
\end{equation}
So to obtain $b(r)$, Eq. (\ref{E:diff3})
must be solved for $\psi^2(r)$.  To
satisfy the condition $b(r_0)=r_0$, we
must have $\psi^2(r_0)=0$, which becomes
the initial condition for Eq.
(\ref{E:diff3}).  Moreover, since
$1-3\psi^2/a^2>0$ by Eq. (\ref{E:diff2}),
we also have $1-\psi^2/a^2>0$, so that
$b(r)>0$.

To check the flare-out condition
$b'(r_0)<1$, we obtain from Eqs.
(\ref{E:shape1}) and (\ref{E:diff3}),
\begin{equation*}
  \left.b'(r)=1-\frac{\psi^2}{a^2}-
  r\frac{(\psi^2)'}{a^2}\right|_{r=r_0}
  =A(8\pi)^{\alpha +1}r_0^{2(\alpha +1)}<1.
\end{equation*}
So to meet the flare-out condition, the
constant $A$ from the Chaplygin equation of
state must satisfy the inequality
\begin{equation}\label{E:A}
   A<\frac{1}{(8\pi r_0^2)^{\alpha +1}},
\end{equation}
which agrees with Refs. \cite{fL06, pK08}.

Now we turn our attention to Eq.
(\ref{E:diff3}), recalling the initial
condition $\psi^2(r_0)=0$.  Since this
equation does not have a closed-form
solution, we will use a numerical approach.
To do so, we choose an arbitrary value for
$r=r_0$ and some values of $A$ and $\alpha$
that satisfy the above conditions.
It becomes apparent immediately that the
solution for $b(r)$ is independent of the
integration constant $a$.  The reason for
this can be seen by writing
Eqs. (\ref{E:diff3}) and (\ref{E:shape1})
in the following respective forms:
\begin{equation}\label{E:diff4}
   \left[\left(\frac{\psi}{a}\right)^2\right]'
   =\frac{1}{r}
   \left[1-\left(\frac{\psi}{a}\right)^2
   -\left(\frac{A(8\pi)^{\alpha +1}
   r^{2(\alpha +1)}}{1-3(\psi/a)^2}
   \right)^{1/\alpha}\right]
\end{equation}
and
\begin{equation}
   b(r)=r\left[1-\left(\frac{\psi}{a}
   \right)^2\right].
\end{equation}
So the solutions of Eqs. (\ref{E:diff3}) and
(\ref{E:diff4}) have the same qualitative forms.
(In other words, by rescaling $\psi$, the
constant $a$ could be eliminated.)

The plots for $\psi^2$ and $b(r)$, using some
typical values of the parameters, are shown in
Fig. 1.  For any particular choice of $A$, the
\begin{figure}[tbp]
\begin{center}
\includegraphics[width=0.8\textwidth]{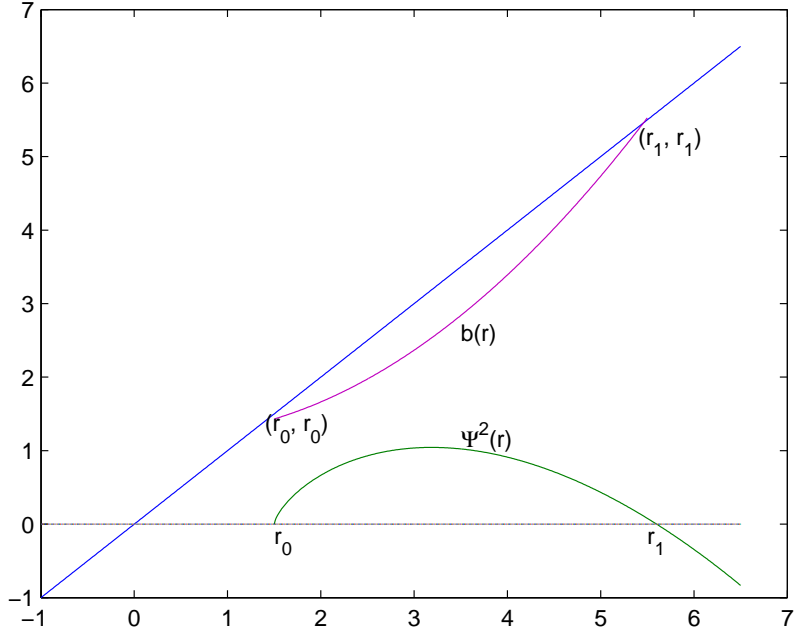}
\end{center}
\caption{$\psi^2$ and $b$ are defined on the
     interval $[r_0, r_1]$.}
\end{figure}
function $\psi^2$ becomes zero for some $r=r_1$ and
hence negative for $r>r_1$.  Since the corresponding
function $\psi$ is now imaginary, $b=b(r)$ is
undefined for $r>r_1$.  (Observe that $b(r_1)=r_1$,
as shown in Fig. 1; also, $b(r)<r$ for $r_0<r<r_1$.)

We conclude that the wormhole material is confined
to some interval $[r_0, r_1]$.  This conclusion is
consistent with Refs. \cite{fL06, pK08}, which state
that the dimensions of the wormhole cannot be
arbitrarily large.

As a final comment, while the length of the interval
$[r_0, r_1]$ is independent of $a$, it does depend
on $A$: the closer $A$ is to the upper limit
$1/(8\pi r_0^2)^{\alpha +1}$, the closer $b'(r_0)$
is to unity and hence the smaller the interval
$[r_0, r_1]$; this behavior can also be seen
from Fig. 1.
%END OF SECTION

\section{The redshift function}

We see from Eq. (\ref{E:gtt}), $e^{\nu}=Cr^2$,
that the wormhole spacetime cannot be asymptotically
flat.  Normally one would now introduce an extra
requirement by stating that the wormhole material
has to be cut off at some $r$ and joined to an
exterior Schwarzschild solution
\begin{equation}
ds^{2}=-\left(1-\frac{2M}{r}\right)dt^{2}
+\frac{dr^2}{1-2M/r}
+r^{2}(d\theta^{2}+\text{sin}^{2}\theta\,
d\phi^{2}).
\end{equation}
Since the need for a cut-off is already known from
the previous section, we are not dealing with a
new requirement.  Unfortunately, the previous
cut-off at $r=r_1$ cannot be used here because
$b(r_1)=r_1$.  Instead, we need to choose some
$r=r_2<r_1$.  We can therefore complete the
wormhole solution by noting that
\[
     M=\frac{1}{2}b(r_2).
\]
So for $e^{\nu}=Cr_2^2$, we have $Cr_2^2
=1-2M/r_2$ and the integration constant becomes
\[
   C=\frac{1}{r_2^2}
   \left(1-\frac{b(r_2)}{r_2}\right).
\]

The need for a cut-off at $r=r_2<r_1$, due to the
coordinate singularity at $r=r_1$, is consistent
with Refs. \cite{vG08, vG09}, which consist of
a detailed investigation of the
Tolman-Oppenheimer-Volkoff equations for
Chaplygin and generalized Chaplygin gas,
respectively.  The main conclusion in these
studies is much more general in the sense that
the scalar curvature was found to become singular
at some finite distance; moreover, the Universe
is not asymptotically flat.  That our wormhole
spacetime cannot be arbitrarily large was also
shown in Ref. \cite{fL06} without relying on
conformal symmetry.  However, in the present
study, conformal symmetry is needed to
determine the redshift function (to obtain 
a complete wormhole solution) but not for the 
wormhole solutions in Refs. \cite{vG08, vG09}.
%END OF SECTION

\section{Conclusion}

For the theoretical construction of a traversable
wormhole, Morris and Thorne proposed the following
strategy: retain complete control over the geometry
by choosing functions $\Phi(r)$ and $b(r)$ that
yield the desired properties, thereby leaving the
components of the stress-energy tensor unspecified.
In this paper we obtained a complete wormhole
solution by (1) adopting the equation of state
$p_r=-A/\rho^{\alpha}$ representing generalized
Chaplygin gas and (2) assuming that the wormhole
admits a one-parameter group of conformal motions.
These two assumptions complement each other via
the requirement that the wormhole spacetime must
be cut off at some $r=r_2$ and joined to an
exterior vacuum solution.

\end{document}